\documentstyle[aps]{revtex}
\input{epsfig.sty}
\def\g5{\gamma^5}

\def\d4k{{d^4k\over (2\pi)^4}}

\def\d{\rm{d}}

\newcommand{\beq}{\begin{eqnarray}}
\newcommand{\eeq}{\end{eqnarray}}
\begin{document}
\title{Instanton Model of QCD Phase Transition Bubble Walls}

\author{Leonard S. Kisslinger\\
        Department of Physics,\\
       Carnegie Mellon University, Pittsburgh, PA 15213}

\maketitle
\indent
\begin{abstract}
Using the liquid instanton model continued into Minkowski space and
modified for finite temperature, the energy momentum tensor and the
surface tension for buble walls during the QCD phase transition are
derived. The resulting surface tension of bubble walls is in agreement 
with lattice calculations. Application to bubble collisions is discussed.
\end{abstract}

\vspace{0.5 in}

\noindent
PACS Indices:12.38.Lg,12.38.Mh,98.80.Cq,98.80Hw

\section{Introduction}
\hspace{.5cm}

   At the time 10$^{-5}$ s the temperature (T) of the universe was 
approximately 300 MeV.  During the time interval between 
10$^{-5}$-10$^{-4}$s the universe passed through the temperature 
T$_c \simeq$ 150 MeV, the critical temperature for the chiral phase 
transition from the quark-gluon plasma (QGP) to the hadronic phase (HP),
the quark-hadron phase transition (QHT). There have been many numerical 
studies of the QHT. Some lattice gauge calculations indicate the the
transition is weakly first order\cite{lat1}, which would imply bubble 
formation during the QHT. At the present time lattice calculations cannot
determine the order of the transition\cite{lat2}. Recent numerical 
calculations in the MIT Bag model\cite{nm} find a first order transition
with hadronic bubbles stable for nucleation at a scale of 1 fm.

  Assuming that the QHT is first-order there have been a number of studies 
of the bubble nucleation\cite{ign} and of possible observational effects of 
such a phase transition. The latter include the possibility of large scale 
density perturbations\cite{is} and the seeding of primary magnetic 
fields\cite{ol} that could lead to observational effects in the 
Cosmic Microwave Background Radiation(CMBR), or large scale galactic 
or extra-galactic structure. 

  The essential property of the bubbles needed to study the nucleation
is the surface tension\cite{ign,ssw}, $\sigma$. The most recent lattice 
calculations give a value of $\sigma \simeq 0.015 \rm{T_c}^3$\cite{bkp}, 
while earlier studies \cite{iw} find a larger value. 
In our present work we wish to estimate
$\sigma$ from the QCD Lagrangian, with the result that we obtain the
energy momentum tensor that can be used in future work on bubble collisions.
For the electroweak phase transition bubble collisions have been studied and
interesting estimates of magnetic fields formed during the collisions have
been made\cite{ae,cst} using an abelian Higgs model plus the QED 
Lagrangian\cite{kib}. However, because of the complexity of nonperturbative
QCD, a reliable model of the bubble walls formed in the QHT has not been
formulated. It is the goal of the present work to formulate such a model.

  Since the main structure of the bubble walls must be gluonic in nature
we use pure gluodynamics. Noting that the walls separate the 
hadronic from the quark/gluon plasma regions of the 
universe, it is natural to consider an instanton picture of the walls.
The QCD instantons are classical solutions for the color field, which
were derived in Euclidean space\cite{bev,hooft}. They connect points
in two vacuua with different winding numbers. In analogy with Coleman's
model\cite{col} of regions of true and false vacuua, we consider the 
instantons connecting vacuua with different winding numbers in the 
quark-gluon region and the hadronic region on the opposite sides of the 
bubble. The Euclidean instanton model cannot be used for the treatment 
of bubble nucleation and collisions (in fact the energy density and
surface tension vanish in this picture), but the model must be continued
to Minkowski space\cite{col}. In brief, our picture is a Minkowski space
analytic continuation of the instanton model for the bubble walls
separating the quark/gluon from the hadronic regions during the QHT.

 It is well known that the instantons of the instanton liquid model\cite{ss} 
can provide the essential nonperturbative gluonic effects of QCD at the medium
length scale of 0.3-0.5 fm, although instantons do not provide confinement.
As we show below, this seems to be the crucial length scale needed for QCD
bubble walls. In recent work an effective Lagrangian was used to estimate 
domain wall formation possibly associated with the
QCD phase transition\cite{zhit}. Although the resulting domain wall is within
the hadronic bubble it resembles the bubble wall that we obtain with the 
instanton model in the present work. That model has also
been used to examine the possibility of magnetic walls formed during the 
QHT seeding gallactic magnetic structure\cite{zhit2}. Such a magnetic
wall has also been considered for cosmic microwave background radiation
polarization correlations\cite{lsk1}. Also, there are
recent numerical investigations of the T-dependence of the properties of the 
QCD instantons\cite{chu,ss2}, which we need in the present investigation.

  In Sec. II we review the instanton liquid model at T=0 and discuss the
energy density and equation of motion in Minkowski space. 
In Sec. III we use recent lattice 
results for instantons at finite temperature\cite{chu} as well as our own 
recent work on glueballs\cite{lsk} to estimate the surface tension.
The surface tension is derived at time t=0, and serves as
an initial condition for bubble nucleation and collisions in Minkowski space.
In Sec. IV we discuss the energy momentum tensor for gluonic QCD in Minkowski
space and possible applications based on the instanton model for
bubble collisions and cosmological observations.

\section{Instanton Model At T=0, Continuation to Minkowski Space, and QCD
 Chirqal Phase Transition Bubbles}
\hspace{.5cm}

   In this section we review the instanton model of QCD at T=0, and the
extension of the model to treat the bubble walls in the QCD chiral phase 
transition. Instantons are obtained and studied in Euclidean space. 
As described in early work on the application of instantons to bubble 
nucleation and collisions\cite{col} in formulating the instanton bubble 
itself one proceeds in Eucledean space, while for the study of collisions 
of nucleating bubbles one must work in Minkowski space. This will be 
discussed further in Sec. IV. 

\subsection{T=0 Instanton Model in Minkowski Space}
The Lagrangian density for pure glue is
\beq
\label{glue}
  {\cal L}^{glue} & = & -\frac{1}{4} G \cdot G,
\eeq
where
\beq
\label{G}
    G_{\mu\nu}^n & = & \partial_\mu A_\nu -  \partial_\nu A_\mu
-i g [A_\mu,A_\nu]\\
    A_\mu & = & A_\mu^n \lambda^n/2 , \nonumber
\eeq
with $\lambda^n$ the eight SU(3) Gell-Mann matrices. The instantons are the
classical solutions of the gauge fields for a pure SU(2) gauge theory.
I.e., writing $A_\mu^n = A_\mu^{n,inst} +A_\mu^{n,qu}$, where 
$A_\mu^{n,inst}$ is a pure gauge classical field, if one keeps only the 
instanton gluonic field , and solves the equation of motion obtained by 
minimizing the classical action in Euclidean space 
\beq
      \delta [ \int G^{inst} \cdot G^{inst}] & = & 0,
\eeq
one obtains the solution\cite{bev}
\beq
\label{inst}
 A_\mu^{n,inst}(x)& = & \frac{2 \eta^{-n}_{\mu\nu}x^\nu}{(x^2 + \rho^2)}
\nonumber\\
        G^{n,inst}_{\mu\nu}(x) & = & -\frac{\eta^{-n}_{\mu\nu} 4 \rho^2}
{(x^2 + \rho^2)^2},
\eeq
for the instanton and a similar expression with -n for the anti-instanton, 
where $\rho$ is the instanton size and the  $\eta^{n}_{\mu\nu}$ 
are\cite{hooft}: $ \eta^a_{bc} = \epsilon_{abc};
\eta^a_{\mu4} = \delta_{a\mu}; \eta^a_{4\nu} = -\delta_{a\nu}$,
with (a,b,c) = (1,2,3) and  $(\mu,\nu)$= (1...4).
The instanton connects points in two QCD 
vacua which differ by one unit of winding number. The instanton contribution 
to the Lagrangian density is
\beq
\label{linst}
  \frac{1}{4} G^{inst} \cdot G^{inst}  & = & 48 \frac{\rho^4}{(x^2+\rho^2)^4}.
\eeq

The energy momentum tensor, $T^{\mu\nu}$, can be obtained from the
Lagrangian, and in Minkowski space is given by (see, e.g., Ref\cite{iz})
\beq
\label{emt}
T^{\mu \nu} &=& \sum_a(G^{\mu \alpha}_a G_{\alpha a}^\nu
-\frac{1}{4}g^{\mu \nu} G^{\alpha \beta}_a G_{\alpha \beta a}).
\eeq
To get the Euclidean space expression for the energy density, $T^{44}$,
one carries out the analytic continuation giving $x^2 = x^\mu x_\mu$ =
${\vec x}^2 +x_4^2$ in Euclidean space, while it is ${\vec x}^2 -x_0^2$ in 
Minkowski space. From this one see that $g^{44}= -g^{00}$. Therefore in
Euclidean space the energy density in the instanton model is
\beq
\label{eucliden}
T_{44}^{inst} & = & G_{a 4 \alpha} G_{a \alpha 4} +\frac{1}{4}G_{a\alpha\beta}
  G_{a\alpha\beta} \nonumber \\
      & = & 0.
\eeq
The result shown in Eq.(\ref{eucliden}) follows from Eq.(\ref{inst}) and
the properties of the  $\eta^{n}_{\mu\nu}$ given in Ref\cite{hooft}.

For the nucleation of QCD bubles and for bubble collisions one must
work in Minkowski space. From Eqs.(\ref{emt},\ref{inst}) one finds for
the energy density in Minkowski space
\beq
\label{minen}
  T^{00,inst}  & = & \frac{1}{2} G^{inst} \cdot G^{inst}
\nonumber\\
     & = & 96 \frac{\rho^4}{(x^2+\rho^2)^4}
\eeq

  Eq.(\ref{minen}) gives the energy density for an instanton. To complete
the instanton model for $T^{00}$ one must find the instanton density, n, for
the particular system in question. Evaluation of the instanton density
is reviewed for both T=0 and finite T in Ref\cite{ss}. In the instanton
model the density is easily evaluated in terms of the gluon condensate, 
which is proportional to the vacuum matrix element $<G \cdot G>$. Thus the 
instanton vlaue for n, n(inst), is given by
\beq
\label{density}
      n(inst) & = & \frac{<G \cdot G>}{\int d^4x  G^{inst} \cdot G^{inst}}.
\eeq
From Eq.(\ref{inst}) and the phenonomenological value for the quark
constant $<G \cdot G> \simeq 1.0 GeV^4$, one finds that
\beq
\label{density1}
        n(inst) & \simeq & 1.0 fm^{-4}.
\eeq
This value is consistent with that determined by the tunneling 
amplitude\cite{hooft,ss}.
Another evaluation of n can be obtained from the QCD sum rule evaluation
of scalar glueballs. Since the correlator for scalar glueballs is given
by the Fourier transform of $<G(x)\cdot G(x) G(0)\cdot G(0)>$, the pole term
in the dispersion relation for the correlator has a factor of n. By comparing
to the 1.5 GeV candidate for a scalar glueball, the best fits were 
found\cite{lsk} for values up to about twice the instanton value. Also, some 
lattice calculations\cite{chu} have found larger values.

To evaluate the surface tension of the bubble walls we must also  consider 
instantons at finite temperature during bubble nucleation, which we do in 
the next section.

\subsection{QCD Chiral Phase Transition Bubble Walls}

The equations of motion for pure glue QCD are obtained from the Lagrangian
given by Eq.(\ref{glue}). Using
\beq
      \delta [ \int G \cdot G] & = & 0,
\eeq
and working in a SU(2) version of QCD, with $A_\mu = A_\mu^a \sigma^a/2$,
$\sigma^a$ being the Paili spin matrices, and $ [A_\mu,A_\nu] = 
i\epsilon_{abc} A_\mu^b A_\nu^c \sigma^a/2$, the equations of motion are
\beq
\label{eom}
\partial_\mu\partial^\mu A^a_\nu 
+ g\epsilon^{abc}(2 A^b_\mu\partial^\mu A^c_\nu 
- A^b_\mu \partial_\nu A^{\mu c}) 
+ g^2\epsilon^{abc}\epsilon^{cef}A^b_\mu A^{\mu e}A^f_\nu & = & 0.
\eeq 
Recognizing that the instanton form of Eq.(\ref{inst}) is a classical
solution to these equations of motion, we try the form
\beq
\label{inst1}
     A_\mu^{n,inst}(x)& = & 2 \eta^{-n}_{\mu\nu}x^\nu F(x^2),
\eeq
and sustituting in Eq.(\ref{eom}) find that $F(x^2)$ satisfies
\beq
\label{eom1}
\partial^2 F &=& -4\frac{\partial F}{\partial x^2}
 - 12 F^2  + 8 x^2 F^3
\eeq
in the Lorentz gauge, with the gauge condition $\partial_\mu A^a_\mu=0$.
By using the initial condition
\beq
\label{initial}
     F(x,t=0) & = & \frac{1}{({\vec x}-{\vec x}_0)^2 + \rho^2},
\eeq
one finds that the instanton form is indeed an approximate solution.
This is discussed further in Section IV, where collisions of QCD
bubbles are discussed.

   From the phenomenological applications\cite{ss} the instanton liquid
model was developed, with $\rho$ = 0.33 fm. This gives the thickness of
the instanton wall at temperature T=0. To complete our model we need the 
instanton density at T=T$_c$, as well s the value of the instanton size 
at that temperature. This is discussed in the next section.

\section{Bubble Wall surface tension}
\hspace{.5cm}

   In this section we evaluate the surface tension in the instanton model.
In order to do this we must consider the dependence of the instanton 
size and density on the temperature. We start with calculating the 
surface tension for an instanton wall. 

   The surface tension is the essential parameter for classical bubble
nucleation. For the treatment of nucleation one must work in Minkowski space. 
As discussed in detail in Ref.\cite{col}, to treat nucleation and collisions 
with instanton input one makes an analytic continuation from Euclidean to 
Minkowski space at the initial time, t=0. For our problem this is achieved
by replacing t by it, or in Minkowski space  $x_\mu x^\mu = {\vec x}^2 - t^2$.
In the present section we study the surface tension at t=0; and the
energy density in Minikowski space derived in the previous section from 
the instanton solutions can be used for the derivation of the surface tension.
The results at t=0 which we find will serve as part of
the initial conditions for the study of nucleation and collisions,
as discussed in Sec. IV. 

The surface energy is the surface tension $\times$ the area of the 
wall for a thin wall. Note that the wall thickness is of the order of 
$\rho$, the instanton size, which is a fraction of a fermi, so that a 
thin-wall formalism is appropriate. Therefore the surface tension with one 
instanton at t=0 is
\beq
\label{iwall1}
       \sigma^{inst} & = & \int dx  T^{00,inst}(x,0,0) \nonumber \\
                     & = & \int dx \frac{1}{2}{\bf G \cdot G}.
\eeq
 From Eqs.(\ref{linst},\ref{iwall1})
\beq
\label{iwall}
  \sigma^{inst} & = & 3\cdot2^5\ \ \int dx \frac{\rho^4}
 {(x^2+\rho^2)^4} \nonumber \\
               & = & 6\cdot2^5\ \ \frac{5!!}{6!!}\frac{\pi}{\rho^3}.
\eeq

   In order to complete our model, we must work with instantons at the
temperature T = T$_c$, which means using a modified instanton size and
density. Moreover, since we are dealing with a gluonic wall separating
the two phases, the QGP and the HP, the number density is quite different
from the result given in Eqs.(\ref{density},\ref{density1}) for the T=0 
system. We use the model applied for the lattice 
calculations\cite{bkp}, and described in detail in Ref.\cite{iw}. In this
picture the free energy is calculated using two peaks for the two phases
on the two sides of the bubble, separated by a mixed quark/gluon-hadronic 
phase. There will be a factor of the instanton number density, i.e., the 
instanton density times the effective four-volume, for each peak.
Therefore, from from Eq.(\ref{iwall}) the resulting surface tension in our 
instanton model is
\beq
\label{sigma}
 \sigma & = & 6\cdot5\ \ \frac{\pi \overline{N}^2}{\overline{\rho}^3},
\eeq
where $\overline{N}$ is the instanton number density at the surface and
$\overline{\rho}$ is the instanton size at T = T$_c$.
We obtain $\overline{N}$ from the T = 0 instanton number per four-volume, 
n = (N/V) of Eq.(\ref{density}). Defining 
n = $\overline{n}$ GeV$^4$ at finite T and using V= $\overline{\rho}^4$, 
$\overline{N}^2$ = 5.96 $\overline{n}^2$, where we use 
$\overline{\rho}$ = 0.25 fm, as discussed next.

At T=0 $\rho$ = 0.33 fm in the liquid instanton model\cite{ss}. This has
also been found in lattice calculations\cite{neg}.
There have been a number of numerical investigations on the modification of
the instanton system at finite T.  Based on these calculation\cite{chu,ss2} 
we take $\overline{\rho}$ = 0.25 fm. There is uncertainty in the value of 
the instanton density, n, at T=0.  In our work on scalar 
mesons/glueballs\cite{lsk} we found values of $\overline{n}$ = 0.0008-0.0015, 
with 0.0008 used in the instanton liquid model. These values also cover
the range of values of lattice gauge calculaitons over the past decade.
From the results of the finite T calculations the n seems to decrease by 
about a factor of 2 at T$_c$, so we use the range 
$\overline{n}$ = 0.0004-0.00075. This gives us the result
\beq
             \sigma & = & (0.013 \rightarrow 0.046) T_c^3,
\eeq
compared to the value $\sigma =$  0.015 T$_c^3$ in the most recent lattice
calculation\cite{bkp}. We note that the instanton liquid model of the QCD 
bubble wall gives a value $\sigma =$  0.013 T$_c^3$ in very good agreement 
with the lattice calculation, and that our model for the bubble wall is 
quite promising.

\section{Discussion of Applications to Collisions}
\hspace{.5cm}

  Although the knowledge of the surface tension is essential for studing
the nucleation and collisions of bubbles during the QCD chiral phase
transition, there is no observable directly related to these bubbles,
so the challenge is to find observable effects related to these bubbles.
In our recent work we have investigated the possibility of a magnetic
walls being formed by interactions of hadrons with an internal gluon 
instanton wall which might be formed within the hadronic phase during
the QCD chiral phase transition. In this section we discuss this possibility
and the relationship to the wall surface tension that has been derived above.

   As was discussed in Sec. II the instanton model itself describes a static
bubble.  For the study of bubble nucleation and collisions one works in 
Minkowski space. Starting from the instanton Lagrangian and using standard
field theory methods one obtains the color ${\bf E}$
and  ${\bf B}$ fields,
\beq
         E^n_i & = &  G^n_{i0} \nonumber \\
         B^n_i & = &  -\frac{1}{2} \epsilon^{ijk} G^n_{jk},
\eeq
with the i,j,k indices running 1,2,3. From this the energy-momentum
tensor, ${\bf T}$ for an instanton in Minkowski space is given as\cite{iz}
\beq
\label{00}
      T^{00,inst} & = & \frac{1}{2}({\bf E}^n\cdot{\bf E}^n + 
 {\bf B}^n\cdot{\bf B}^n) \\
\label{01}     
      T^{0i,inst} & = & \epsilon^{ijk} E^n_j B^n_k \\
\label{ij}      
      T^{ij,inst} & = & (E^{ni}E^{nj} +B^{ni}B^{nj}) -T^{00,inst}\,\delta_{ij}
\eeq
  From the results of Sec. II and with a straightforward calculation we obtain
for the stress energy tensor in the instanton model from Eq.(\ref{ij})
\beq
  T^{ij,inst} & = & - \, \left( \frac{4 \overline{\rho}^2}
{(x^2 +\overline{\rho}^2)^2} \right)^2 \, \delta_{ij} .
\eeq
To obtain the color  ${\bf E}$ abd  ${\bf B}$ fields one must solve the
QCd equations of motion.  Solutions to the equations of motion with an
effective Lagrangian have been found for a pure SU(2) 
gauge theory in Minkowski space\cite{fks}. Using the general picture
of Coleman\cite{col} in which one sets the instanton initial conditions
at time t=0 and then uses the equations of motion in Minkowski space for 
the evolution, a SU(2) model of of QCD instantons has been developed and
applied to heavy relativistic heavy ion collisions\cite{ocs}. 

   We are developing a similar program for our cosmological studies of the 
QCD chiral phase transition in the early universe based on QCD field theory.
During the collision of two bubbles the colliding walls might form an
internal wall within the merged bubbles. In classical nucleation theory
when two identical bubbles collide they merge to form a bubble with an 
interior wall having the same surface tension as the original bubble walls.
In our case the result would be instanton walls described in the previous
sections. Note that in Ref.\cite{zhit} an estimate of the
lifetime of the gluonic domain walls in their model was 10$^{-5}$s, which
allows magnetic walls to be formed\cite{zhit2}. Although the present theory
is quite different, a two-dimensional model of bubble collisions starting
from the gluonic QCD Lagrangian finds that an internal gluonic wall is
formed\cite{jck}, and this suggests that magnetic structures could form.
Using the standard evolution methods as used for
tangled magnetic fields for the scattering \cite{sub} or for
metric perturbations\cite{dfk} one can investigate possible observational
effects in the CMBR. A preliminary investigation of this has been carried 
out\cite{lsk1} assuming the interior instanton wall is formed, which is
shown to lead to an interior magnetic wall, and CMBR polarization 
correlations are found that could be observed in the next few years. 

   The crucial question is whether an internal wall is formed with the
properties of the nucleating QCD bubbles. In a 1+1 dimensional study in 
pure QCD, the equations of motion  of Eq.(\ref{eom1}) were solved. This
corresponds physically to very large bubbles colliding, as illustrated 
in Fig.1.
\begin{figure}
\centerline{\psfig{figure=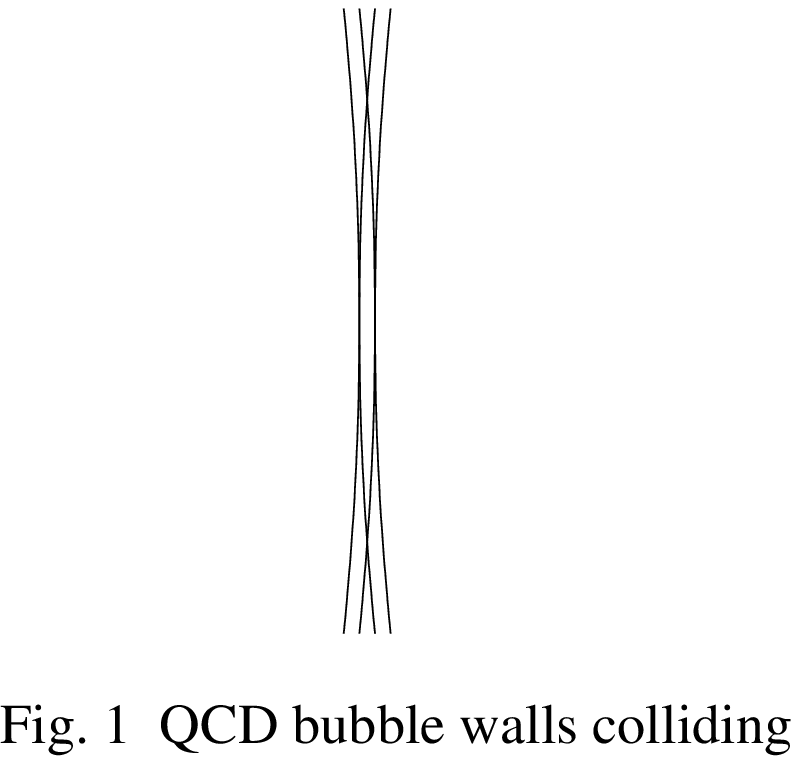,height=5cm,width=12cm}}
{\label{Fig.1}}
\end{figure}
To model collisions one takes an initial condition with two walls,
such as
\beq
\label{initial1}
  F(x,t=0) & = & \frac{1}{( x-x_0)^2 +\rho^2}+\frac{1}{( x+x_0)^2 +\rho^2},
\eeq
with $2x_0$ the separation between the two bubble walls at time t=0,
and follows the evolution. In Ref\cite{jck} it was found that an interior
wall is indeed formed, which seems to be similar to the initial walls. Fig.1
illustrates the process at the instant when the walls are overlapping,
and the calculation shows the two bubbles separating with an internal
instanton-like structure remaining at the collision region.
From this one would conclude that the interior wall would have approximately
the same surface tension as discussed in the previous sections, and 
could result in the magnetic wall of Ref\cite{lsk1}. Moreover, the studies
of the QCD chiral phase transition\cite{ign,is}  find that in
contrast to the electroweak phase transition where many bubles nucleate,
the QCD chiral phase transition seems to proceed via inhomogeneous 
nucleation, with larger distance between bubbles and few nucleating bubbles 
involved in the transition to the hadronic phase. Thus the large bubble
collision of Fig.1 and Ref\cite{jck} is well-founded.

  To investigate the nucleation and collision of bubbles during the QCD
chiral phase transition one must include quark and hadronic degrees of 
freedom, since they play a major role in the free energy difference 
between the hadronic and quark/gluon phases. With Eq.(\ref{eom}), 
however, one can learn a great deal about the bubble walls, and our
results for the surface tension in the present paper are quite promising.
The full 3+1 dimensional treatment of nucleating QCD bubbles is a
subject of future research.

\vspace{1mm}
{\bf Acknowledgements}\\
The author would like to acknowledge helpful discussions with Mikkel Johnson, 
Ernest Henley, Pauchy Hwang and Ho-Meoyng Choi.
This work was supported in part by NSF grant PHY-00070888 and by the Taiwan
CosPA Project, Taiwan Ministry of Education 89-N-FA01-1-3

\end{document}